\documentclass[aps, twocolumn,floats, showpacs]{revtex4}
\usepackage{amsmath,amssymb,graphicx}
\usepackage{epsfig}
\usepackage{graphicx}
\usepackage{enumerate}

\newcommand{\avg}[1]{\left< #1 \right>} 
\renewcommand{\d}[2]{\frac{d #1}{d #2}} 

\let\baraccent=\= 
\newcommand{\intinf}{\int_{-\infty}^\infty}

\newcommand{\beq}{\begin{equation}}
\newcommand{\eeq}{\end{equation}}

\newcommand{\bea}{\begin{eqnarray}}
\newcommand{\eea}{\end{eqnarray}}

\begin{document}


\title{Quantum fluctuation theorem for heat exchange in
the strong coupling regime}
\author{Lena Nicolin}
\affiliation{Chemical Physics Theory Group, Department of Chemistry, University of Toronto,
80 Saint George St. Toronto, Ontario, Canada M5S 3H6}
\author{Dvira Segal}
\affiliation{Chemical Physics Theory Group, Department of Chemistry,
University of Toronto, 80 Saint George St. Toronto, Ontario, Canada
M5S 3H6}

\date{\today}
\begin{abstract}
We study quantum heat exchange in a multi-state
impurity coupled to two thermal reservoirs. Allowing for strong
system-bath interactions, we show that a steady-state heat exchange
fluctuation theorem holds, though the dynamical processes
nonlinearly involve the two reservoirs.
We accomplish a closed
expression for the cumulant generating function, and use it obtain
the heat current and its cumulants in a nonlinear thermal junction, the two-bath
spin-boson model.
%
\end{abstract}


\maketitle


Exact fluctuation relations for nonequilibrium classical systems have
been recently discovered and exemplified, dealing with work and entropy fluctuations
in various (open, closed, driven) systems \cite{FT1}.
In particular, the fluctuation theorem (FT) for entropy production quantifies
the probability of negative entropy generation, measuring "second law violation" \cite{Evans,Cohen}.
Both transient and steady-state fluctuation theorems (SSFT) have been derived,
where the latter measures entropy production in nonequilibrium
steady-state systems over a long interval. In the context of heat
exchange between two equilibrium reservoirs, $\nu=L, R$, the SSFT
can be roughly stated as  $\ln [P_{t}(+\omega)/P_t(-\omega)]=\Delta
\beta \omega$ \cite{Jarzynski, Abe}. Here $P_t(\omega)$ denotes the probability
distribution of the net heat transfer $\omega$, from $L$ to $R$,
over the (long) interval $t$, with $\Delta \beta=T_R^{-1}-T_L^{-1}$
as the difference between the inverse temperatures of the
reservoirs. A related quantity is the cumulant
generating function (CGF), providing general relations between
transport coefficients under the FT symmetry \cite{Mukamel-rev,Saito}.

Extending the work and heat FT to the quantum domain has recently
attracted significant attention \cite{Mukamel-rev, Hanggi-rev}.
Specifically, a quantum exchange FT, for the transfer of energy
between two reservoirs maintained at different temperatures, has
been derived in Refs. \cite{Jarzynski, Tasaki, Campisi} using
projective measurements, and in Refs. \cite{Mukamel-weak,
HanggiBerry}, based on the unraveling of the quantum master equation
(QME). These derivations assume that the interaction between the two
thermal baths is weak, and can be neglected with respect to overall
energy changes. Using the Keldysh approach, an exact analysis was
carried out in \cite{Dhar}. However, it is valid only for harmonic
systems. It is thus an open question whether a heat exchange FT is
obeyed by an anharmonic quantum system {\it strongly coupled} to
multiple reservoirs.

From a practical point of view, understanding and controlling energy
transport and heat dissipation in nanoscale junctions is crucial for
making further progress in device miniaturization \cite{Pop}.
Theoretical studies adopting simple models can reveal the role of
different system parameters on the transport mechanisms
\cite{Segalheat,Rectif,Galperin,Thoss}. However, such treatments
either assume weak coupling between the nanoscale object and the
environment, an assumption that is not always justified, or are
limited to very simple models.

It is our objective here to investigate quantum heat exchange in
two-terminal impurity models: (i) To derive the SSFT for heat
currents in open quantum systems, incorporating anharmonic
interactions, allowing for strong system-bath interactions ("strong
coupling"). (ii) To obtain the CGF and gain explicit expressions for
the heat current and its second moment, useful for understanding
heat current characteristics for anharmonic-strongly coupled
systems. (iii) To understand the role of nonmarkovian (memory)
effects on the onset of the  SSFT.

Our analysis begins with a general model for the impurity,
reservoirs and the interaction form. Describing the dynamics at the
level of the noninteracting-blip approximation (NIBA) \cite{Weiss},
a scheme accommodating strong system-bath interactions, we derive a
QME for the system dynamics, under the markovian limit. Unraveling
these equations into trajectories with a particular amount of net
energy dissipated, e.g., to the $R$ reservoir, a heat exchange SSFT
is verified. We also obtain the CGF, independent of the particular
physical realization. The scheme is exemplified on the two-terminal
spin-boson model. In the {\it nonmarkovian} case a general symmetry
relation is recovered, whereas the universal SSFT is reached in the
markovian limit only.


{\it Model.---} Consider a quantum impurity (system)
placed between two thermal reservoirs (baths). No
assumptions are made on the energy structure of the impurity, thus anharmonic
systems, with finite and uneven energy spacings, are comprised.
Further, system-bath interactions are potentially strong relative to
the system energetics.
We adopt the dressed-tunneling Hamiltonian,
\bea H&=&\sum_n \epsilon_n |n\rangle \langle n| +\sum_{\nu}H_{\nu}
\nonumber\\ &+& \sum_{n>m}\frac{\Delta_{nm}}{2}\left(|n\rangle
\langle m| e^{-i\Omega_{nm}} + |m\rangle \langle n| e^{i\Omega_{nm}}
\right), \label{eq:HG} \eea
where $|n\rangle$ denotes the impurity quantum states, coupled
through the tunneling elements $\Delta_{nm}$, dressed by the baths
operator $\Omega_{nm}=\Omega_{nmL}+\Omega_{nmR}$. The operators
$\Omega_{nm\nu}$ depend on the coordinates of the $\nu=L,R$ bath and
may represent, for example, a collection of displacements or
momentum operators as in the standard small polaron model
\cite{Weiss}. Furthermore, different bath operators may couple to
different transitions. The thermal reservoirs $H_{\nu}$  are assumed
to be in a canonical state, maintained at a temperature
$T_{\nu}=\beta_{\nu}^{-1}$. Besides that, we do not specify the
reservoirs, and they may be composed of fermions, spins, photons or
phonons. The Hamiltonian (\ref{eq:HG}) allows only for energy
transfer processes between the two baths, mediated by a system
excitation. Transfer of particles is not considered in the present
study.

{\it population dynamics.---} System dynamics is explored at the
level of the NIBA scheme \cite{Aslangul,Legget,Dekker}: Applying the
Born approximation \cite{Weiss} to the dressed Hamiltonian
(\ref{eq:HG}), equations of motion for the impurity reduced density
matrix can be readily obtained \cite{Rectif}. This approximation is
generally valid for $\Delta<\omega_c$, where $\omega_c$ is a cutoff
of the reservoirs modes, at high temperatures and in the strong
coupling regime \cite{Weiss}. Neglecting coherences and for
simplicity, further applying the Markov approximation, we get
quantum kinetic equations for the population $p_n$,
\bea \dot p_n=-p_n\sum_{m\neq n} C_{nm}(\omega_{nm})+ \sum_{m\neq n}
p_m C_{nm}(\omega_{mn}). \label{eq:pop} \eea
The transition rate from state $n$ to $m$, $C_{nm}(\omega_{nm})$, is
a convolution of $L$-induced and $R$ induced processes
\cite{Rectif},
\bea C_{nm}(\omega_{nm})&=&\intinf  e^{i\omega_{nm} t} C_{nmL}(t)
C_{nmR}(t)dt
\nonumber\\
&=& \intinf  C_{nmL}(\omega_{nm}-\omega)C_{nmR}(\omega) d\omega.
\label{eq:rate}
\eea
Here $\omega_{nm}=\epsilon_n-\epsilon_m$. The indices of
$C_{nm}$ are ordered such that $n>m$. The $\nu$-bath correlation
function is given by the thermal average
\bea C_{nm\nu}(t)=\frac{\Delta_{nm}}{2}\langle
e^{i\Omega_{nm\nu}(t)} e^{-i\Omega_{nm\nu}(0)}\rangle. \eea
The operators are written in the interaction representation,
$\Omega_{nm\nu}(t)=e^{iH_{\nu}t}\Omega_{nm\nu} e^{-iH_{\nu}t}$. In
frequency domain we write $C_{nm\nu}(\omega)=\intinf dt e^{i\omega
t} C_{nm\nu}(t)$, which are the elements in (\ref{eq:rate}). As a
result of microreversibility, detailed balance is satisfied for each
reservoir, separately,
\bea \frac{C_{nm\nu}(\omega)}{C_{nm\nu}(-\omega)}=
e^{\omega\beta_{\nu}}. \label{eq:DB}
 \eea
Such a detailed balance condition does not hold for the
combined rate $C_{nm}(\omega_{nm})$ since it
encloses both temperatures
through the bath-specific
correlations $C_{nm\nu}$.

While the dynamics is simply described by a QME, it still encloses
complex physical processes. Eq. (\ref{eq:rate}) draws nontrivial
transfer rates. For example, when the system decays making a
transition from state $n$ to $m$, it disposes the energy
$\omega_{nm}$ into both reservoirs cooperatively; an energy $\omega$
is dissipated into the $R$ bath while the $L$ bath gains (or
contributes) the rest, $\omega_{nm}-\omega$. Similarly, excitation
of the system occurs through an $L$-$R$ compound process. We
highlight the three non trivial mechanisms involved here, arising
due to the strong coupling limit: (i) Non-resonance energy transfer
processes are allowed, where each reservoir donates (absorbs) an
energy which does not overlap with the system's energy spacings.
(ii) Anharmonic processes are allowed. For example, in the context
of vibrational energy transfer multiphonon processes are
incorporated within the relaxation rates $C_{nm}$, see e.g., Eq.
(\ref{eq:Q}). (iii)  The transport process takes place conjoining
the reservoirs' dynamics in a non-additive manner, as discussed
above. In contrast, the weak coupling limit, studied in
\cite{Rectif,Mukamel-weak,HanggiBerry} in the context of bosonic
transfer, admits only resonant transmission processes and single
phonon effects. Moreover, in the weak coupling limit the reservoirs
additively act on the system \cite{comm}.

{\it Cumulant Generating function.---} We define the function
$P_t(n,\omega)$ as the probability distribution that within the time
$t$ a net energy $\omega$ has been dissipated into the $R$ bath,
with the system populating the $n$ state at time $t$. For later use
we also construct $P_t(\omega)=\sum_n P_t(n,\omega)$,  the
distribution of $\omega$ at $t$, irrespective of the system state.
The time evolution of $P_t(n,\omega)$ obeys
\bea &&\dot P_{t}(n,\omega)=
 \sum_{m\neq n} \intinf \Big[ P_t(m,\tilde \omega)C_{nmR}(\omega-\tilde \omega)
\nonumber\\
&&\times C_{nmL}(\tilde \omega-\omega-\omega_{nm}) d\tilde \omega
\Big]
\nonumber
\\
&&-P_t(n,\omega)
\sum_{m\neq n} \intinf C_{nmR}(\tilde \omega)
C_{nmL}(\omega_{nm}-\tilde \omega) d\tilde \omega.
\label{eq:PRw}
\eea
This can be justified by energy-resolving the population dynamics in
(\ref{eq:pop}), then collecting the matching energy terms from the
left side and the right side of the equation. The first term here
describes a process where by the time $t$ a net energy $\tilde
\omega$ has been damped into $R$, whereas the system  occupies the
state $m$. At the moment $t$ the system (assisted by the bath)
transits from $m\rightarrow n$, further dissipating an energy
$\omega-\tilde\omega$ into the $R$ reservoir.
Similarly, the second term collects all transitions which deplete $P_t(n,\omega)$.
Next we introduce the counting field $\chi$ and Fourier transform
the resolved probabilities,
$P_t(n,\chi)=\intinf d\omega e^{i\omega\chi} P_t(n,\omega)$,
yielding
\bea &&\dot P_t(n,\chi)= -P_t(n,\chi)\sum_{m\neq n}
C_{nm}(\omega_{nm}) \nonumber\\
&&+ \sum_{m> n} P_t(m,\chi) f_{mn}^+(\chi)+\sum_{m<n} P_t(m,\chi)
f_{nm}^-(\chi). \label{eq:PR} \eea
For brevity, we introduce the short notation
\bea f_{nm}^{\pm}(\chi)= \intinf e^{i\omega \chi}
C_{nmR}(\omega)C_{nmL}(\pm\omega_{nm}-\omega) d\omega.
\label{eq:f}
\eea
%
These equations can be encapsulated in a matrix form
%
$|\dot \Psi(\chi,t)\rangle=-\hat \mu(\chi) |\Psi(\chi,t)\rangle$,
%
with $\Psi$ a vector of the probabilities $P_t(n,\chi)$.  We define
the characteristic function  $Z(\chi,t)=\langle I| \Psi(\chi,t)
\rangle$, with $\langle I|$, as a left vector of unity, and the {\it
cumulant generating function}
%
$G(\chi) =  \lim_{t \to \infty} \ \frac{1}{t}\ln Z(\chi,t)$,
%
recovered as the negative of the smallest eigenvalue of the matrix $\hat \mu$.

{\it Steady-state fluctuation theorem.---} We now prove that $G(\chi)=G(i\Delta
\beta-\chi)$, implying that a SSFT for heat exchange holds.
In order to derive this relation we analyze the symmetry properties of
the matrix $\hat \mu$. For clarity, we explicitly write it
for a three-state impurity
\bea \hat \mu(\chi)= \left( \begin{array}{ccc}
 \mu_{1,1}  & -f_{21}^{+}(\chi) & -f_{31}^+(\chi)  \\
-f_{21}^-(\chi) &  \mu_{2,2}  & -f_{32}^+(\chi)  \\
-f_{31}^-(\chi) & -f_{32}^-(\chi) & \mu_{3,3} \\
\end{array} \right)
\eea
The diagonal terms $\mu_{i,i}$ constitute the decay rates from each level, and
are  independent of $\chi$. The characteristic polynomial $D_{\hat
\mu(\chi)}(\lambda)$, with the roots $\lambda$, is given by
\bea &&D_{\hat\mu(\chi)}(\lambda)=
f_{31}^+(\chi)\left[f_{21}^-(\chi)f_{32}^-(\chi) -
(\lambda-\mu_{2,2}) f_{31}^-(\chi)\right]
\nonumber\\
&&-f_{21}^+(\chi)\left[f_{21}^-(\chi)(\lambda-\mu_{3,3}) -
f_{32}^+(\chi)f_{31}^-(\chi)\right]
\nonumber\\
&&+(\lambda-\mu_{1,1}) \left[(\lambda-\mu_{2,2})(\lambda-\mu_{3,3})-
f_{32}^-(\chi)f_{32}^+(\chi)\right].
\nonumber \eea
%
One can show that the following three properties hold: (i)  $\hat
\mu(\chi)$ is symmetric under the operation
$f_{nm}^+(\chi)\rightarrow f_{nm}^-(\chi)$. 
Thus, the roots $\lambda$  are also symmetric in this
respect.
%
(ii) Each element in the characteristic polynomial is cyclic, in the
sense that a series of transitions must end at the initial state.
For example, the product $f_{31}^+(\chi)f_{21}^-(\chi)
f_{32}^-(\chi)$ describes a relaxation process from state 3 to 1,
followed by an excitation from state 1 to 2, finishing with an
excitation term $f_{32}^-(\chi)$, bringing the system back to state
3. (iii) The correlation function $f_{nm}^+(\chi)$ satisfies the
identity
%
%
\bea
f_{nm}^+(i\Delta \beta - \chi)= e^{\beta_L\omega_{nm}} f_{nm}^-(\chi),
\label{eq:fs}
\eea
gathered by manipulating Eq. (\ref{eq:f}) with (\ref{eq:DB}).
%
%
Under these three properties we prove that
$D_{\hat \mu(\chi)}(\lambda)=D_{\hat\mu(i\Delta
\beta -\chi)}(\lambda)$: The symmetric terms in the characteristic polynomial
are mapped one onto the other as a result of the symmetry (\ref{eq:fs}) whereas
the {\it system dependent} prefactors, i.e.,
the term $e^{\beta_L\omega_{nm}}$ in Eq. (\ref{eq:fs}) overall
cancel, a result of the cyclic property (ii). We
conclude that the eigenvalues of $\hat \mu$ satisfy a
symmetry relation, and in particular $G(\chi)=G(i\Delta
\beta-\chi)$. The probability distribution of $\omega$ is obtained as
$P_t(\omega)=\frac{1}{2\pi} \intinf d\chi Z(\chi,t)e^{-i\chi
\omega}$.
Since $Z(\chi,t)\sim e^{G(\chi)t}$ in the long time limit,
a heat exchange fluctuation relation is resolved
\bea
\lim_{t\rightarrow
\infty}\frac{1}{t}\ln \frac{P_t(\omega)}{P_t(-\omega)} =\frac{\Delta
\beta \omega}{t}.
\eea
We emphasize: This relation has been derived without specifying
neither the system energy structure and its interaction with the
reservoirs, nor the form of the reservoirs. It allows for
strong coupling between the impurity and the baths, reflected in the
transition rates $C_{nm}$, mixing $L$-$R$ processes in a
non-additive manner. Moreover, an explicit expression for the CGF, $G(\chi)$,
can now be written, bearing analytical expressions for the current
cumulants, as we achieve below for the spin-boson model.

{\it Spin boson model.---}
The equilibrium spin-boson (SB) model, referring to  a
spin immersed in an equilibrated boson reservoir, is an eminent
model in chemistry and physics, useful for describing, e.g., solvent
assisted electron transfer reactions and the Kondo resonance
\cite{Weiss}. The nonequilibrium spin-boson model, where the spin is
coupled to more than one thermal reservoir, has been suggested as a
prototype model for exploring heat transfer through nanojunctions
\cite{Rectif,Thoss}.
We now analytically obtain the CGF, thus the current and its
moments, for the nonequilibrium SB model at strong coupling,
\beq
H = \frac{\omega_0}{2} \sigma_z +  \frac{\Delta}{2}\sigma_x +
\sigma_z \sum_{\nu,j}\lambda_{j,\nu}(b_{j,\nu}^\dagger + b_{j,\nu}) +
\sum_{\nu,j}\omega_j b_{j,\nu}^{\dagger}b_{j,\nu}.
\label{eq:HSB}
\eeq
Here $\sigma_x$ and $\sigma_z$ are the Pauli matrices, $\omega_0$ is
the energy gap between the spin levels, and $\Delta$ is the
tunneling energy.
The two reservoirs include a collection of uncoupled harmonic oscillators,
$b_{j,\nu}^{\dagger}$ ($b_{j,\nu}$) is the bosonic creation
(annihilation) operator of the mode $j$ in the $\nu$ reservoir. The
parameter $\lambda_{j,\nu}$ accounts for the system-bath interaction
strength. The Hamiltonian is transformed to the
displaced bath-oscillators basis using the small polaron
transformation \cite{Weiss}, $H_S=U^{\dagger}HU$,
$U=e^{i\sigma_z\Omega/2}$,
%
\bea
H_{S} = \frac{\omega_0}{2} \sigma_z +
\frac{\Delta}{2} \left( \sigma_+ e^{i\Omega} + \sigma_- e^{-i\Omega} \right)
+\sum_{\nu,j}\omega_j b_{j,\nu}^{\dagger}b_{j,\nu},
\label{eq:HSBs}
\eea
where $\sigma_{\pm}=\frac{1}{2}(\sigma_x\pm i \sigma_y)$
are the auxiliary Pauli matrices, $\Omega=\sum_{\nu}\Omega_{\nu}$,
and
$\Omega_{\nu}=2i\sum_{j}\frac{\lambda_{j,\nu}}{\omega_{j}}(b_{j,\nu}^{\dagger}-b_{j,\nu})$.
Under the NIBA, the system population obeys a convolution-type
master equation \cite{Aslangul,Legget,Dekker} ($\langle \sigma_z
\rangle =p_1-p_0$),
%
\bea
 \dot p_1=
-\frac{\Delta^2}{2} \int_{0}^{t}
e^{-Q'(t-s)}
\cos[ \omega_0(t-s)-Q''(t-s)]
p_1(s) ds
 \nonumber\\
 +\frac{\Delta^2}{2} \int_{0}^{t} e^{-Q'(t-s)} \cos[
\omega_0(t-s)+Q''(t-s)] p_0(s)ds,
\nonumber
\label{eq:integrP}
\eea
%
with conserved total occupation $p_0(t)+p_1(t)=1 $. The function
$Q(t)=\sum_{\nu}{Q_{\nu}(t)}$, made of a real and imaginary
components, $Q_{\nu}(t)=Q'_{\nu}(t)+iQ''_{\nu}(t)$, is defined by
\bea
Q'_{\nu}(t)& = & \int_{0}^{\infty}\frac{J_{\nu}(\omega)}{\pi\omega^2}[1-\cos(\omega t)] [1+2n_{\nu}(\omega)] d\omega,
\nonumber\\
Q''_{\nu}(t)& = &  \int_{0}^{\infty} \frac{J_{\nu}(\omega)}{\pi\omega^2}\sin(\omega t)d\omega.
\label{eq:Q}
\eea
Here  $J_{\nu}(\omega)=4\pi\sum_{j}\lambda_{j,\nu}^2\delta(\omega-\omega_j)$
is the $\nu$-bath spectral function, $n_{\nu}(\omega)$ is the Bose-Einstein distribution.
We have carried out the analysis in the {\it nonmarkovian} limit by
generalizing Eq. (\ref{eq:PRw}), to describe the dynamics of
$P_t(n,\omega_L,\omega_R)$, for the transfer of $\omega_{\nu}$ net
energy to the $\nu$ bath by the time $t$. Introducing two counting
fields $\chi_{1,2}$, then following the procedure outlined in Ref.
\cite{Antti} (applying Fourier transform and Laplace transform on
the resolved equation of motion, analyzing the poles of the
resolvent), we can  prove that the CGF satisfies \cite{Future}
\bea
G(\chi_1,\chi_2)=G(i\beta_L-\chi_1,i\beta_R-\chi_2).
\label{eq:Gn}
\eea
Only in the markovian limit the symmetry is given in terms of the affinity as
$G(\chi)=G(i\Delta \beta-\chi)$.
Thus, while microreversibility is sufficient for deriving
the basic symmetry relation (\ref{eq:Gn}), the SSFT holds only under more restrictive conditions,
dictated here by the bath relaxation timescale \cite{Mukamel-rev,Tasaki}.

In the markovian case
the QME for the population dynamics [Eqs. (\ref{eq:pop})-(\ref{eq:rate})] follows
$\dot p_1=-C(\omega_0)p_1 +C(-\omega_{0})p_0$,
with the rates $C(\omega_0) =
\intinf e^{i\omega_0 t} C_L(t) C_R(t) dt$; $C_{\nu}(t)=
e^{-Q_{\nu}(t)}$.
Since only a single correlation function matters,
 the level indices were discarded. Following Eqs.
(\ref{eq:PRw})-(\ref{eq:f}), we identify the matrix $\hat \mu$ by
\bea \hat{\mu}(\chi) =
\begin{pmatrix}
C(-\omega_0) & -f^+(\chi)  \\
-f^-(\chi) & C(\omega_0)  \\
\end{pmatrix}
\label{eq:mu}
\eea
with $f^{\pm}(\chi)=\intinf
e^{i\omega\chi}C_R(\omega)C_L(\pm\omega_0-\omega)d\omega$. Its smallest
eigenvalue is
\bea
&&G(\chi)= -\frac{1}{2}[C(\omega_0)+C(-\omega_0)]
\nonumber\\
&&+ \frac{1}{2}\sqrt{  (C(\omega_0)-C(-\omega_0))^2+4f^-(\chi)f^+(\chi) }.
\label{eq:G}
\eea
%
\begin{figure}[htbp]
\vspace{0mm} \hspace{0mm} {\hbox{\epsfxsize=65mm \epsffile{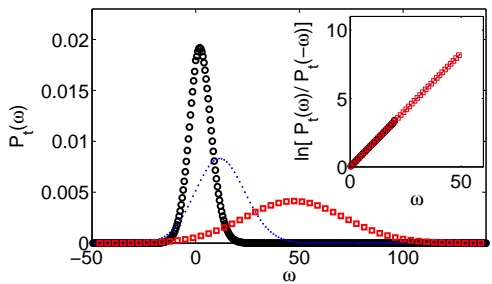}}}
\caption{Nonequilibrium spin-boson model: Plot of $P_{t}(\omega)$ at
various times. The inset demonstrates the validity of the SSFT.
$T_L=3$, $T_R=2$, $E_r^{\nu}=1$, $\omega_0=0.5$, $t=20$ ($\circ$),
$t=100$ (dotted) and $t=400$ ($\square$). } \label{Fig1}
\end{figure}
The averaged heat current can be readily obtained,
%
\bea
&&\langle J\rangle \equiv \frac{\avg \omega_t }{t}
= \d{G(\chi)}{(i\chi)}\Big|_{\chi=0}=
\int_{-\infty}^{\infty}
\Big[ C_R(\omega)C_L(\omega_0-\omega)p_1
\nonumber\\
&&- C_R(-\omega)C_L(-\omega_0+\omega)p_0 \Big]\omega d\omega.
\label{eq:J}
\eea
The population here is calculated in steady-state, $p_0=C(\omega_0)/[C(\omega_0)+ C(-\omega_0)]$.
This expression was heuristically suggested in Ref. \cite{Rectif},
and here it is derived from the basic dynamics.
Note that the details of the function $Q(t)$
are not utilized in this derivation.
Furthermore, the averaged current stays intact for nonmarkovian systems \cite{Antti}.
The formal structure for the noise power is given by
\bea
\langle S \rangle &=&
\frac{d^2G(\chi)}{d(i\chi)^2}\Big|_{\chi=0}= - 2\Big[C(\omega_0)+C(-\omega_0)\Big]^{-1}
\nonumber\\
&\times&
\Big[\intinf\omega C_-(\omega) d\omega \intinf\omega C_+(\omega)d\omega + \avg{J}^2 \Big]
\nonumber\\
&+& \intinf d\omega \omega^2 \Big[C_+(\omega) p_1+ C_-(\omega)
p_0\Big], \eea
where we defined $C_{\pm}(\omega)=
C_R(\pm\omega)C_L(\pm\omega_0\mp\omega)$.
We can also plot the distribution $P_t(\omega)$. Assuming high
temperatures $T_{\nu}>\omega_0$ and strong coupling, Eq.
(\ref{eq:Q}) can be simplified, $Q'_{\nu}(t)=E_r^{\nu}T_{\nu}t^2$,
$Q_{\nu}''(t)=E_r^{\nu} t$, with the reorganization energy defined
as $E_r^{\nu}=\sum_{j}4\lambda_{j,\nu}^2/\omega_j$  \cite{Marcus}.
Using this form, Fig. \ref{Fig1} displays the entropy production
distribution and the validity of the SSFT (inset).

To conclude, a heat exchange  SSFT has been derived for quantum
systems incorporating strong system-bath interactions and anharmonic
effects.
Our study provides closed expressions for the CGF, useful for
deriving the distribution of heat fluctuations, the averaged current
and the thermal noise power. For the spin-boson model one can show
that in the nonmarkovian case the SSFT does not generally hold. It
is satisfied in the markovian limit, when energy conservation is
enforced.
Future work will be devoted to generalizing our study to systems
showing coherence effects.

LN is funded by the Early Research Award of DS.
Support from NSERC is acknowledged.



\begin{thebibliography}{25}
\bibitem{comm}
In the weak coupling limit we expand
$e^{i\Omega}=e^{i\Omega_L}e^{i\Omega_R}\sim 1+i\Omega_L+i\Omega_R$ in Eq. (\ref{eq:HG}).
Following the Born-Markov approximation with the system-bath coupling strength as the small parameter,
one finds that the system relaxation rate is additive with respect to the two reservoirs
\cite{Rectif}.
\end{thebibliography}
\end{document}